# Imaging spontaneous currents in superconducting arrays of π-junctions

Sergey M. Frolov[1], Micah J.A. Stoutimore[1], Dale J. Van Harlingen[1], Vladimir A. Oboznov[2], Valery V. Ryazanov[2], Adele Ruosi[3], Carmine Granata[3], Maurizio Russo[3]

[1]*Department of Physics, University of Illinois at Urbana-Champaign, 1110 W. Green St., Urbana, IL 61801, USA;* [2]*Institute of Solid State Physics, Russian Academy of Sciences Chernogolovka, 142432, Russia;* [3]*Instituto di Cibernetica "E. Caianiello" del CNR Via Campi Flegrei 34 Comprensorio A. Olivetti, Pozzuoli, Napoli, Italy*

**Two superconductors separated by a thin tunneling barrier exhibit the Josephson effect and can be characterized by a single macroscopic wavefunction that allows charge transport at zero voltage[1], typically with no phase shift between the superconductors in the lowest energy state. Recently, Josephson junctions with the ground state phase shifts of π proposed by theory three decades ago[2], have been demonstrated[3–5]. Enclosed in superconducting loops, π-junctions cause spontaneous circulation of persistent currents[2], making such loops analogous to spin-1/2 systems[6]. Here we use a scanning SQUID (Superconducting QUantum Interference Device) microscope[7] to image spontaneous currents in superconducting networks of temperature-controlled π-junctions based on weakly ferromagnetic barriers[3]. By combining cells with even and odd numbers of π-junctions, we construct non-uniformly frustrated arrays that have previously not been attained. We find an onset of spontaneous supercurrents at the 0-π transition temperature of the junctions $T_\pi \sim 3$ K. Control over both geometry and interaction strength in these arrays makes them attractive as model systems for studies of the exotic phases of the 2D XY-model[8-9] and for applications in scalable architectures of adiabatic quantum computers[10].**

The ground state wavefunction of a conventional Josephson junction has the same phase in both superconductors and supports a supercurrent in the direction of the applied phase difference $\phi$, as described by the classic Josephson current-phase relation $I_s(\phi) = I_c \sin(\phi)$, where $I_c > 0$ is the critical current. In a $\pi$-Josephson junction, it is energetically favorable for the superconducting ground state wavefunction to change its sign across the tunneling barrier so that a small phase difference drives a supercurrent in the opposite direction. Thus, $\pi$-Josephson junctions have a negative critical current $I_c < 0$ and a current-phase relation of the form: $I_s(\phi) = -|I_c| \sin(\phi) = |I_c| \sin(\phi + \pi)$. The Josephson coupling can be described by the overlap of the superconducting wavefunctions that decay exponentially into the barrier from the superconducting electrodes, a phenomenon known as the proximity effect. If the barrier is a ferromagnet, the exchange interaction introduces an energy splitting between the spin-up and spin-down electrons forming the Cooper pairs, leading to a finite center-of-mass momentum state for the pairs[11,12]. This results in a damped spatial oscillation of the proximity-induced superconducting wavefunction[13], which strongly modifies the Josephson effect. In fact, such oscillations can be directly observed in the dependence of the critical current of superconductor-ferromagnet-superconductor (SFS) junctions on the ferromagnet thickness[14]. As the barrier thickness is varied, the coupling between the wavefunctions of the left and right superconductors oscillates and changes sign, inducing a series of transitions into and out of the $\pi$-junction state. For this experiment, we prepared arrays of Nb-$Cu_{47}Ni_{53}$-Nb junctions near the first (thinnest) 0-$\pi$ transition barrier thickness, approximately d = 11 nm for the weak ferromagnetic alloy used. Since the ferromagnetic barrier is diffusive, the wavelength of the order parameter oscillation depends on temperature. In our junctions, the oscillation period changes by ~ 0.1 nm per degree K, allowing transitions between 0-junction and $\pi$-junction states to be observed in a single junction as a function of temperature.

A remarkable manifestation of the π-state of Josephson junctions is the generation of spontaneous persistent currents in superconducting loops incorporating odd numbers of π-junctions. We call such currents spontaneous because no applied magnetic fields or power sources are required to create or sustain them. These currents arise in order to satisfy the fluxoid quantization conditions in the loops in response to the π-shifts across the junctions. Because the ground state of a loop with a single π-junction is doubly-degenerate with respect to the spontaneous current direction, π-junctions have been proposed as elements for superconducting flux qubits[10] and for rapid single flux quantum (RSFQ) logic circuits[15]. We fabricated two-dimensional square arrays of SFS Josephson junctions with various dimensions. In each array, some of the cells were frustrated with three π-junctions, while other cells had either two or four π-junctions and were unfrustrated (Fig.1b). In the unfrustrated cells, the ground state configuration corresponds to each junction being in its lowest energy state with a phase drop of π and there is no circulating current. In contrast, frustrated cells (with sufficient inductance) require a spontaneous current to maintain flux quantization and energy minimization conditions. Therefore, while screening currents may circulate in all cells in the presence of applied magnetic fields, spontaneous currents only appear in frustrated cells.

A photograph of a 6x6 uniformly frustrated array (all cells frustrated) is shown in Fig.1c. To elucidate the structure of the array, a sketch of a single frustrated cell is shown in Fig.1a. Each cell is created by overlapping two pairs of cross-shaped superconducting Nb electrodes of thickness 100 nm (base) and 240 nm (top). An insulating layer of 100 nm thick SiO with windows of area $4 \times 4$ μm$^2$ created by lift-off lithography separates the two superconducting layers. In three of the four overlap areas, ferromagnetic $Cu_{47}Ni_{53}$ barriers of thickness d = 11 nm are deposited to create SFS Josephson junctions. In the fourth overlap area, the base and the top superconductors are connected directly to form a superconducting short instead of an SFS junction. Unfrustrated cells would have ferromagnetic barriers in all four (or two of four) overlap areas. The periodicity of our arrays is 30 μm, with each cell having an open area of 15 × 15 μm$^2$ corresponding to a geometric inductance of ~ 25 pH.

The Scanning SQUID Microscope (SSM) measures the average vertical magnetic field in a superconductor pickup loop scanned over the surface of a planar sample. The pickup loop, which is coupled to a dc SQUID detector via a superconducting flux transformer, is fabricated on a Si wafer that is bevelled to form a tip. The sensor assembly is hinged so that it rests at a small angle (~ 5 deg) from the substrate with the tip in contact with the surface, maintaining the pickup coil at a distance of about 3 μm from the surface. The SSM has a spatial resolution of 5-10 μm, determined by the size of the pickup loop, and a magnetic flux sensitivity of $10^{-5}$ $\Phi_0$, where $\Phi_0 \approx 2.05 \times 10^{-15}$ Wb is the quantum of magnetic flux. SSM imaging has also been used in the past to investigate spontaneous currents in loops incorporating d-wave superconductors[16], in which intrinsic phase shifts of π between orthogonal tunneling directions[17] cause similar phase frustration.

Images of unfrustrated arrays showed virtually no contrast in zero magnetic field at all temperatures. However, in a small magnetic field, below the threshold field at which vortices enter the array, the array structure becomes visible because the superconducting islands screen the inductance of the SQUID pickup coil, as shown in Fig.1d for a 6×6 cell array. In contrast, all arrays with frustrated cells exhibit spontaneous circulating currents. As an example, Fig.1e shows a scanning SQUID microscope image of a uniformly frustrated 6×6 array cooled below the π-state crossover in zero magnetic field. Over most of the array, the arrangement of spontaneous currents is antiferromagnetic, with the direction of the spontaneous currents alternating in adjacent array cells to produce a checkerboard magnetic flux pattern, as expected for the ground state configuration. However, deviations from antiferromagnetic patterns are often observed, including in this image, since arrays can cool into metastable excited states in which one or more spontaneous currents are flipped.

Antiferromagnetic current patterns observed in a uniformly frustrated array cannot serve as definitive proof of spontaneous currents and of the existence of a π-junction state. Indeed, a magnetic flux pattern like the one shown in Fig.1e can occur due to screening currents in an array of regular junctions if the applied magnetic flux is close to $0.5\Phi_0$ per cell. As a result, previous phase-sensitive experiments, while capable of demonstrating temperature-driven 0-π transitions, relied on the knowledge of the order parameter oscillation length to define the π-junction side of the transition[18,19]. In order to distinguish unambiguously spontaneous currents induced by π-junctions from screening currents that arise in superconducting loops when magnetic fields are present, we have prepared arrays with non-uniform frustration. In Fig.2a we show an SSM image of a variety of such arrays obtained at T = 1.5 K, along with the corresponding frustration patterns indicated by the diagrams. In all of our images taken in nominally zero applied magnetic field, currents only occur in the frustrated cells, a unique signature of the π-junction state. Fig.2b shows two 3×3 arrays imaged in small but finite magnetic field B ≈ 2 mG. The effect of the magnetic field is to induce all of the spontaneous currents in the same direction.

Close to the transition into the π-junction state, the critical current density of SFS junctions is highly sensitive to variations in the ferromagnetic barrier thickness (at the rate of 100 A/cm$^2$ per 1 Å) or the ferromagnet exchange interaction. All of the frustrated cells we have imaged have the same area and contain three nominally identical SFS junctions so that the magnitude of flux in each cell should be comparable. However, as illustrated by Fig.2c, which shows a one-dimensional uniformly-frustrated array, the magnitudes of the spontaneous currents vary from cell to cell by as much as 50% due to variations in the critical currents of SFS junctions. From variations in spontaneous currents we estimate that at T = 1.5 K, the base temperature of the SSM, the critical current density fluctuates in the range 100-400 A/cm$^2$ in our junctions, consistent with the effective barrier thickness variations of 1-5 Å previously observed in such junctions[20].

The energies of spontaneous current configurations can be obtained from simulations of the Josephson phase dynamics of the arrays. In the simulations, the phase differences across the junctions are randomly initiated, and then allowed to evolve according to the equations describing the Josephson dynamics and the loop phase coherence until relaxed into either the ground state or a metastable state corresponding to an irregular arrangement of spontaneous flux. The energy spectra of the π-junction arrays along with the SSM images of the ground and the first metastable states are given in Fig.3 for fully frustrated and checkerboard-frustrated (alternating frustrated and unfrustrated cells) 2×2 arrays. It can be noted that, in contrast to arrays of conventional 0-junctions, the lowest energy state is achieved at a finite value of the applied magnetic flux. When the applied magnetic flux is zero, the lowest energy state is antiferromagnetic for both types of frustration. However, the energy gap separating the ground state from the first metastable state, with one of the spontaneous currents flipped, is lower in checkerboard-frustrated arrays. This is because the closest current-carrying cells do not share a branch but rather couple via adjacent unfrustrated cells. The direct magnetic coupling of currents is negligible in our arrays. It can also be seen from the simulations that the Josephson energy never reaches zero in checkerboard-frustrated arrays as it does in fully frustrated arrays. This is true for any non-uniform frustration pattern and indicates that in the π-junction state currents circulate in non-uniformly frustrated arrays at any applied magnetic field.

We can monitor the onset of spontaneous currents in an array by taking images at a series of different temperatures. We expect such currents to onset only below the π-state transition temperature $T_\pi$ of the individual junctions in the array. To estimate $T_\pi$, we measured the temperature dependence of the critical current of a single isolated SFS junction fabricated on the same wafer as the arrays. The critical current, obtained from the current-voltage characteristics as shown in Fig.4a, vanishes at $T \approx 2.9K$, which we identify as $T_\pi$.

The nucleation of spontaneous currents in multiply connected circuits incorporating π-junctions depends on the energy balance between the Josephson coupling energies of the junctions and the magnetic field energy in the loops. For a single π-junction in a loop of inductance L, a spontaneous current is induced only if the loop is large enough that the inductance parameter $\beta_L = 2\pi L I_c/\Phi_0 > 1$. In this regime, the inductive energy $E_L = LJ^2/2$ associated with the circulating current J is lower than the Josephson coupling energy $2E_J = I_c\Phi_0/\pi$ which must be paid to keep the π-junction in its highest energy state with a zero phase difference. Adding either more junctions to a single cell or more cells to an array increases the effective inductance and reduces the magnitude of the critical current at which spontaneous currents onset. Thus, in large arrays we expect the onset of spontaneous currents to occur very near the transition of the individual junctions into the π-state.

The temperature evolution of spontaneous currents in a 6×6 checkerboard-frustrated array is shown in Figs.4b-4d. No spontaneous currents were observed at temperatures well above $T_\pi$, as illustrated by the SSM image in Fig.4b, which was obtained at T = 4.0 K and shows no contrast. In Fig.4c, after we cool to T = 2.8 K just below the expected transition into the π-junction state, flux from spontaneous currents can be discerned but individual vortices cannot be resolved. The reason is that the critical currents of SFS junctions become very small close to $T_\pi$ so that $\beta_L \ll 1$, making the characteristic size of the vortices much larger than one cell. In this regime, the spontaneous flux generated by each frustrated cell is much smaller than $0.5\ \Phi_0$ (by roughly a factor of $1/\beta_L$) . We note that in the vicinity of $T_\pi$ temperature becomes a useful experimental knob for studying vortex dynamics because small changes in temperature of order 0.1K-1K result in 2-3 orders of magnitude change in the Josephson coupling energy. At the same time, properties of the superconducting material remain nearly constant, since $T_\pi$ is far from the superconducting critical temperature $T_c$. Fig.4d shows an SSM image of the array obtained at T = 1.6 K. At temperatures well below $T_\pi$ the spontaneous currents are bigger, and the checkerboard-frustration pattern can be resolved. In Fig.4e we plot the rms SQUID voltage obtained by averaging the SSM signal over the array at different temperatures. Fore comparison, on the same graph we show the calculated temperature dependence for the onset of spontaneous magnetization in an isolated cell frustrated by a single π-junction. The onset of spontaneous currents in the array is in the vicinity of T = 3 K, and is broadened compared to the simulation due to variations in the SFS junction barrier thicknesses, which create a spread in 0-π transition temperatures of individual junctions. The ratio of the Josephson coupling energy $E_J$ to the inductive energy $E_L$ required to generate magnetic flux of $0.5\Phi_0$ is $E_J/E_L=(I_c\Phi_0/2\pi)/(\Phi_0^2/8\pi^2 L)\approx 5$ at our lowest achievable temperature T = 1.5 K, meaning that spontaneous magnetic flux in each frustrated cell reaches $0.4\Phi_0$.

1. Josephson, B.D. Possible new effects in superconductive tunneling. *Phys. Lett.* **1**, 251 (1962).

2. Bulaevskii, L.N., Kuzii, V.V. & Sobyanin, A.A. Superconducting system with weak coupling to a current in the ground state. *JETP Lett.* **25**, 290 (1977).

3. Ryazanov, V.V. et al. Coupling of two superconductors through a ferromagnet: evidence for a π-junction. *Phys. Rev. Lett.* **86**, 2427 (2001).

4. Baselmans, J.J.A., Morpurgo, A.F., van Wees, B. & Klapwijk, T.M. Reversing the direction of supercurrent in a controllable Josephson junction. *Nature* **397**, 43 (1999).

Work supported by the National Science Foundation grant EIA-01-21568, by the U. S. Civilian Research and Development Foundation (CRDF) grant RP1-2413-CG-02, and by the Russian Foundation for Basic Research.

Correspondence and requests for materials should be addressed to DVH (dvh@uiuc.edu).

Fig. 1. **π-junction arrays.** (**a**) Schematic of a single frustrated array cell. Ferromagnetic CuNi layers are sandwiched between cross-shaped superconducting Nb electrodes to form SFS junctions. (**b**) In an unfrustrated cell with an even number of π-junctions, the spontaneous flux is zero in the lowest energy state; in a frustrated cell with 3 π-junctions spontaneous currents generate magnetic flux of order $\Phi_0/2$. (**c**) Optical image of a 6×6 fully-frustrated array. (**d**) SSM image of an unfrustrated array in small applied magnetic flux

( $\ll\Phi_0/2$ per cell) showing contrast from superconducting niobium. (**e**) SSM image of a fully-frustrated 6×6 array in the π-state with zero applied field.

Fig. 2. **Designed frustration.** SSM images of arrays in the π-state at T= 1.5 K with diagrams indicating the frustration patterns. (**a**) Zero applied field SSM image of 2×2 arrays of various frustration showing spontaneous currents only in frustrated cells (**b**) Antiferromagnetic state of a fully-frustrated line array showing fluctuations in the magnitude of spontaneous currents (**c**) 3×3 arrays in small applied magnetic field.

Fig. 3. **Spontaneous current configurations.** Calculated Josephson energies as a function of the applied magnetic flux and SSM images of corresponding spontaneous current configurations in 2×2 arrays with (**a**) full frustration and (**b**) checkerboard frustration.

Fig. 4. **Temperature-driven onset of spontaneous currents.** (**a**) Temperature dependence of the critical current of a single SFS junction showing transition in the π-junction state. The sign of the critical current is indicated but cannot be determined from current-voltage characteristics. (**b-d**) SSM images of a 6×6 checkerboard frustrated array taken at T = 4.0K, 2.8K and 1.6K. (**e**) Variance of the magnetic flux generated in the array at different temperatures (solid dots) superimposed with calculated spontaneous flux onset curve for a loop of single inductance L = 25 pH with a single SFS junction (solid line). The dashed line indicates offset due to SQUID detector noise.

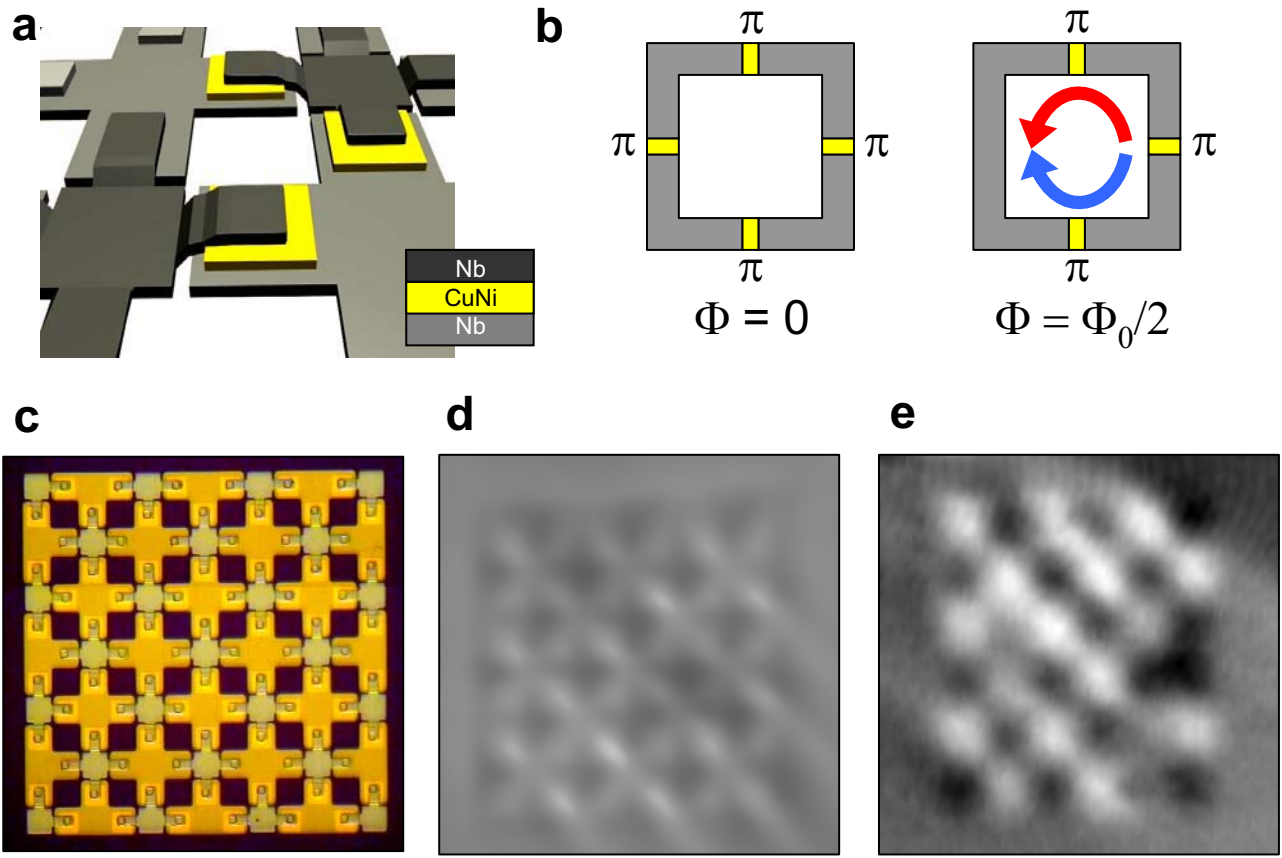

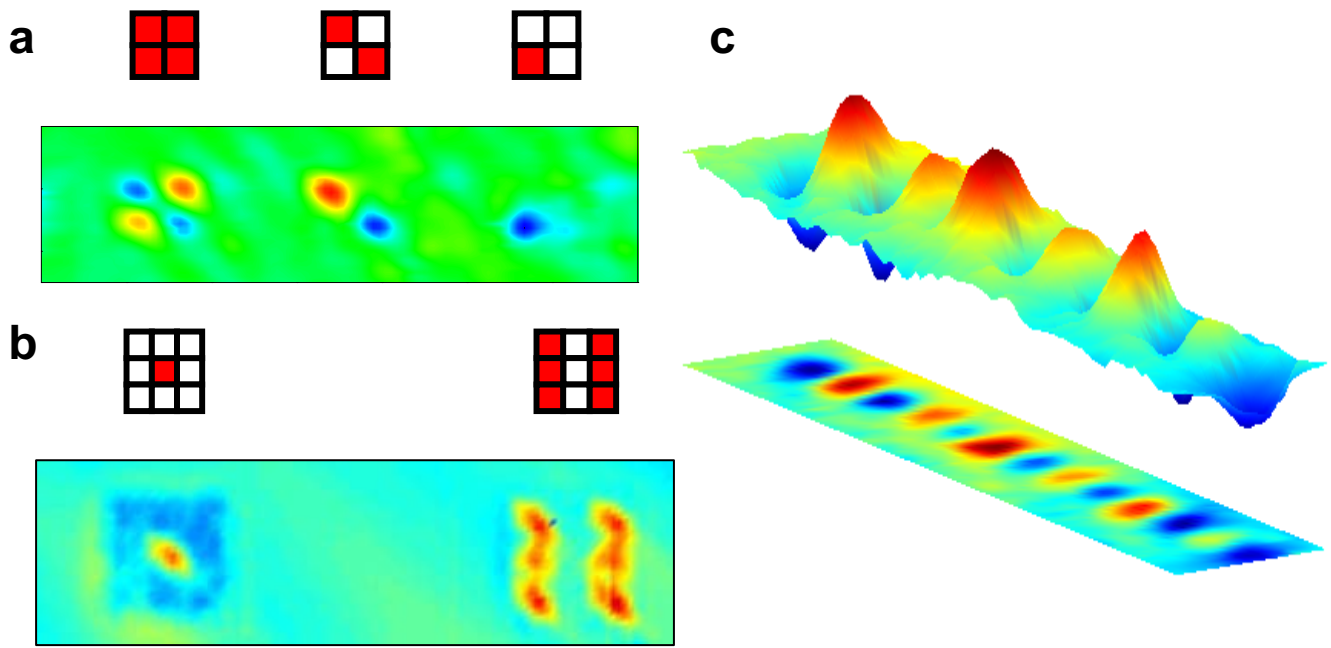

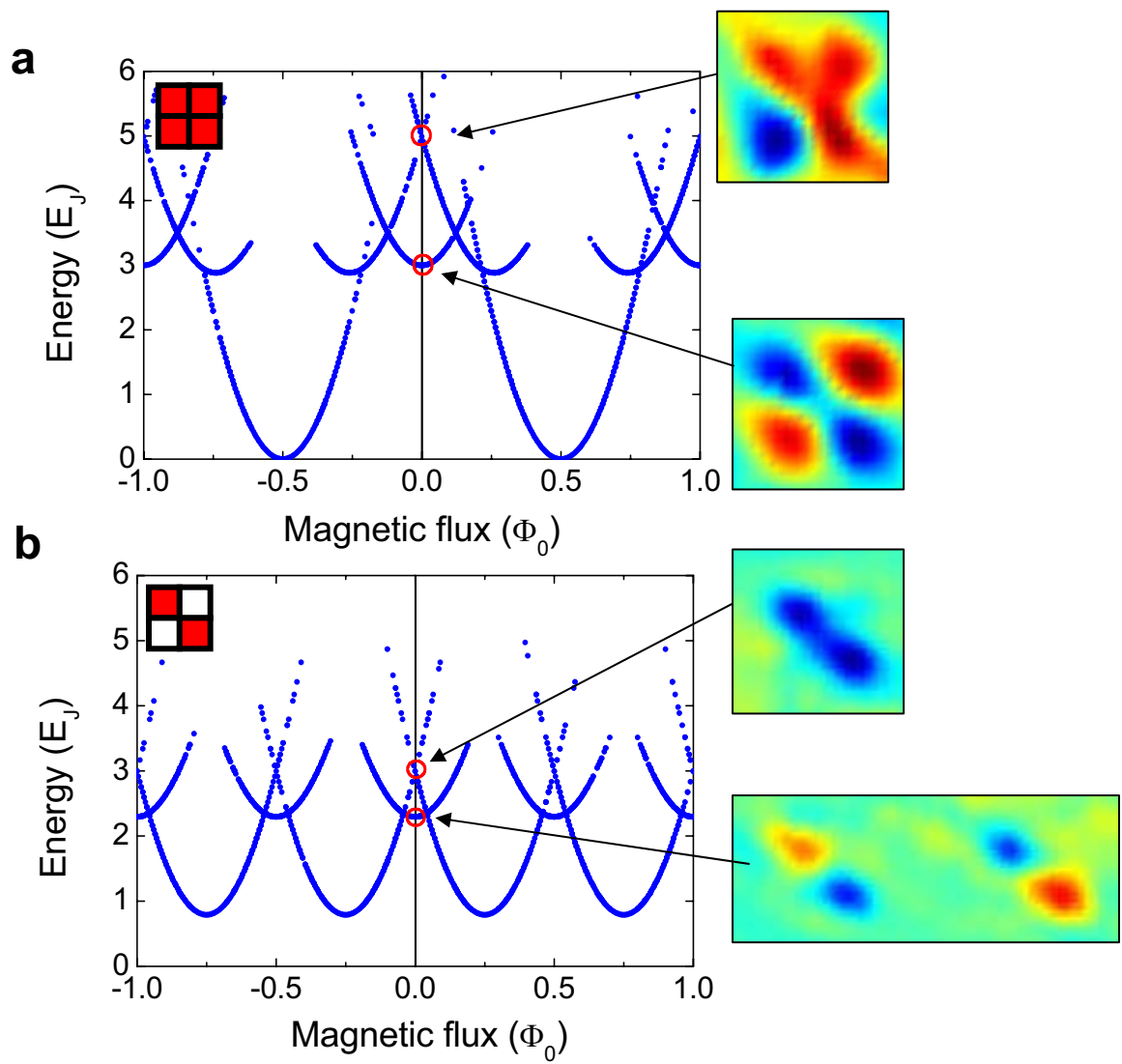

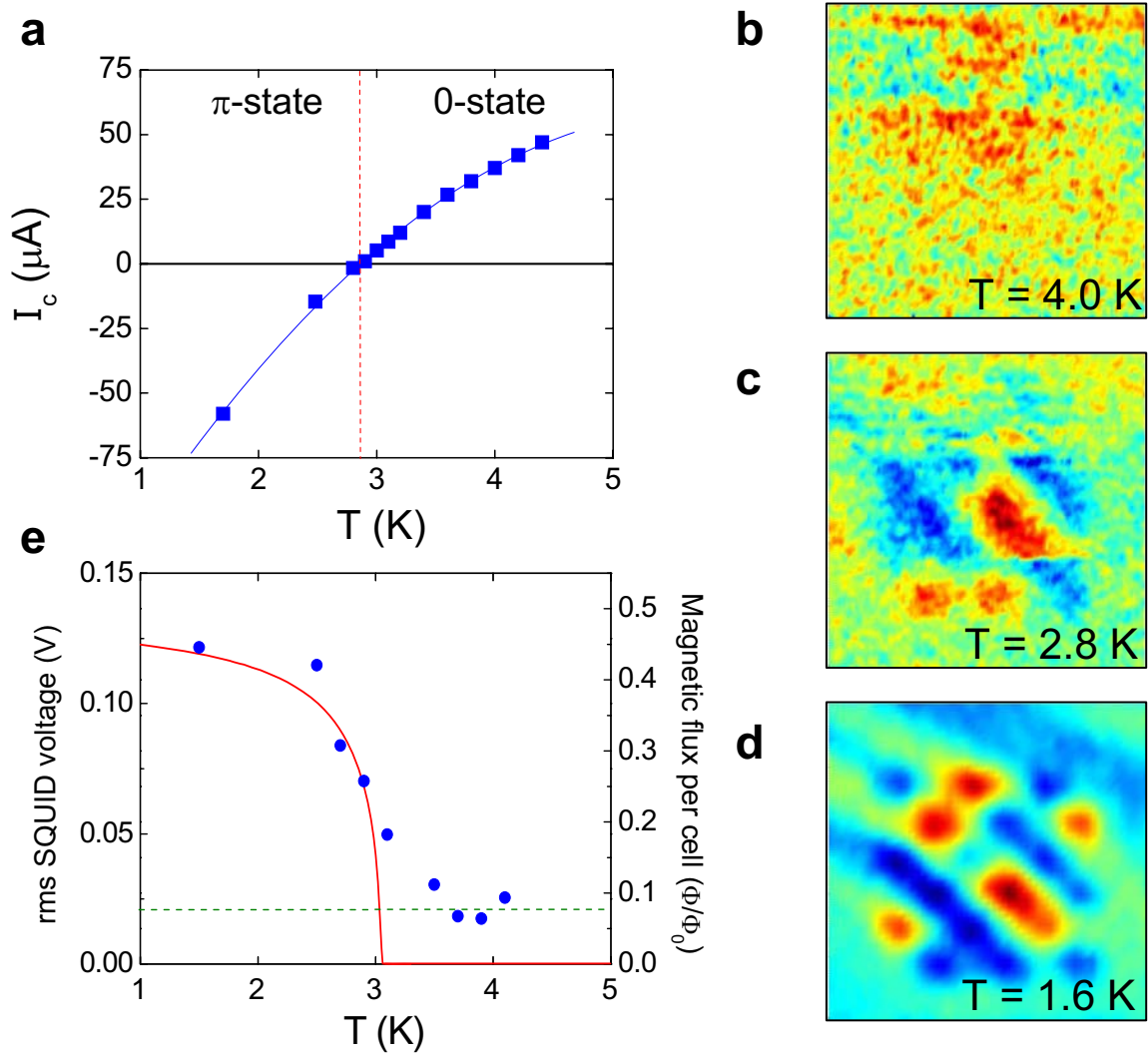

Van Harlingen    Figure 4